\documentclass{nature}
\usepackage{graphicx}
\makeatletter
\let\saved@includegraphics\includegraphics
\AtBeginDocument{\let\includegraphics\saved@includegraphics}

\makeatother

\usepackage{multirow}
\usepackage{subfigure}
\usepackage{mathtools}
\usepackage[colorlinks=true,linkcolor=blue,citecolor=black,urlcolor=blue]{hyperref}
\usepackage{authblk}
\usepackage[labelfont=bf]{caption}
\usepackage[figurename=Figure]{caption}
\usepackage{siunitx}
\usepackage[dvipsnames]{xcolor}
\usepackage{soul}
\usepackage{xfrac}
\usepackage{amsmath}
\usepackage{soul,xcolor}
\usepackage{lineno}
\usepackage{siunitx}

\graphicspath{{./figures/}}

\setstcolor{red}

\title{Emergence of zero-field non-synthetic single and interchained antiferromagnetic skyrmions in thin films}
\author{Amal Aldarawsheh$^{1,2*}$, Imara Lima Fernandes$^1$, Sascha Brinker$^1$, Moritz Sallermann$^{1,3,4}$, Muayad Abusaa$^5$, Stefan Bl\"ugel$^1$ \&  Samir Lounis$^{1,2*}$}
\begin{document}
\maketitle
\begin{affiliations}
 \item Peter Gr\"{u}nberg Institute and Institute for Advanced Simulation, Forschungszentrum J\"{u}lich and JARA, D-52425 J\"{u}lich, Germany
 \item Faculty of Physics, University of Duisburg-Essen and CENIDE, 47053 Duisburg, Germany
 \item RWTH Aachen University, 52056 Aachen, Germany
 \item Science Institute and Faculty of Physical Sciences, University of Iceland, VR-III, 107 Reykjavík, Iceland
 \item Department of Physics, Arab American University, Jenin, Palestine\
 
 * a.aldarawsheh@fz-juelich.de; s.lounis@fz-juelich.de
 
\end{affiliations}

\begin{abstract}
Antiferromagnetic (AFM) skyrmions are envisioned as ideal localized topological magnetic bits in future information technologies. In contrast to ferromagnetic (FM) skyrmions, they are immune to the skyrmion Hall effect, might offer potential terahertz  dynamics while being insensitive to external magnetic fields and dipolar interactions. Although observed in synthetic AFM structures and as complex meronic textures in intrinsic AFM bulk materials, their realization in non-synthetic AFM films, of crucial importance in racetrack concepts, has been elusive. Here, we unveil their presence in a row-wise AFM Cr film deposited on PdFe bilayer grown on fcc Ir(111) surface. Using first principles, we demonstrate the emergence  of single and strikingly interpenetrating chains of AFM skyrmions, which can co-exist with the rich inhomogeneous exchange field, including that of FM skyrmions, hosted by PdFe. Besides the  identification of an ideal platform of materials for intrinsic AFM skyrmions, we anticipate the uncovered knotted solitons to be promising building blocks in AFM spintronics.
\end{abstract}

\section*{Introduction}
Magnetic skyrmions are particle-like topologically protected twisted magnetic textures~\cite{Bogdanov1989,bogdanov1994thermodynamically,Roessler2006} with exquisite and exciting properties~\cite{Nagaosa2013,fert2013skyrmions}. They often result from the competition between the Heisenberg exchange and the relativistic Dzyaloshinskii-Moriya interaction (DMI)~\cite{dzyaloshinsky1958thermodynamic,Moriya1960}, which is present in materials that lack inversion symmetry and have a finite spin orbit coupling. Since their discovery in multiple systems, ranging from bulk, thin films, surfaces to  multilayers~\cite{muhlbauer2009skyrmion,pappas2009chiral,yu2010real,yu2012skyrmion,yu2011near,heinze2011spontaneous,romming2013writing,Gong2015,Soumyanarayanan2017}, skyrmions are  envisioned as promising  candidates for bits, potentially usable in the transmission and storage of information in the next generation of spintronic devices~\cite{kiselev2011chiral,crum2015perpendicular,Wiesendanger2016,Garcia-Sanchez2016,Fert2017,fernandes2020defect}.
However, requirements for future (nano-)technologies are not only limited to the generation of information bits but are also highly stringent from the point of view of simultaneous efficiency in reading, control, and power consumption~\cite{Fert2017,Zhang2020}. Miniaturization of ferromagnetic (FM) skyrmions suffers from the presence of dipolar interactions~\cite{Buettner2018}, while their stabilization generally requires an external magnetic field. Another drawback is the skyrmion Hall effect~\cite{Nagaosa2013}, caused by the Magnus force that deflects FM skyrmions when driven with a current, which hinders the control of their motion. Additionally, FM skyrmions exhibit a rather complex dynamical behavior as function of applied currents~\cite{Lin2013,Woo2016,Jiang2016,Litzius2017,fernandes2018universality,Fernandes2020a,Arjana2020}, under the presence of defects.

Antiferromagnetic (AFM) skyrmions are expected to resolve several of the previous issues and offer various advantages. Indeed, AFM materials being at the heart of the rapidly evolving field of AFM spintronics~\cite{Jungwirth2016,olejnik2018terahertz,gomonay2018antiferromagnetic} are much more ubiquitous than ferromagnets. Their compensated spin structure inherently forbids dipolar interactions, which should allow the stabilization of rather small skyrmions while enhancing their robustness against magnetic perturbations. AFM skyrmions were  predicted early on using continuum models~\cite{bogdanov2002magnetic}, followed with multiple phenomenology-based studies on a plethora of properties and applications, see e.g. Refs.~\cite{rosales2015three,keesman2016skyrmions,zhang2016antiferromagnetic,zhang2016thermally,gobel2017antiferromagnetic,bessarab2019stability,potkina2020skyrmions,liu2020theoretical,shen2018dynamics,silva2019antiferromagnetic,khoshlahni2019ultrafast,Diaz2019,zarzuela2019stabilization}.   The predicted disappearance of the Magnus force, which triggers the skyrmion Hall effect, would then enable a better control of the skyrmion’s motion~\cite{zhang2016antiferromagnetic,barker2016static}, which has been partially illustrated experimentally in a ferrimagnet~\cite{woo2018current,hirata2019vanishing}.

Intrinsic AFM meronic spin-textures (complexes made of half-skyrmions) were recently detected in bulk phases~\cite{Gao2020,Ross2020,Jani2021} while synthetic AFM skyrmions were found within  multilayers~\cite{legrand2020room}. However, the observation of intrinsic AFM skyrmions has so far been  elusive, in particular at surfaces and interfaces, where they are highly desirable for racetrack concepts. A synthetic AFM skyrmion consists of two FM skyrmions realized in two different magnetic layers, which are antiferromagnetically coupled through a non-magnetic spacer layer. In contrast to that an intrinsic AFM skyrmion is a unique magnetic entity since it is entirely located in a single layer. Here we predict from first-principles (see Method section)  intrinsic AFM skyrmions in a monolayer of Cr deposited on a surface known to host ferromagnetic skyrmions: A PdFe bilayer grown on Ir(111) fcc surface as illustrated in Fig.~\ref{fig:1}a. The AFM nature of Cr coupled antiferromagnetically to PdFe remarkably offers the right conditions for the emergence of a rich set of complex AFM textures. The ground state is collinear row-wise AFM (RW-AFM) within the Cr layer (see inset of Fig.~\ref{fig:1}c), a configuration hosted by a triangular lattice so far observed experimentally only in Mn/Re(0001)~\cite{spethmann2020discovery,spethmann2021discovery}. The difference to the latter, however, is that although being collinear, the Cr layer interfaces with a magnetic surface, the highly non-collinear PdFe bilayer.  

A plethora of  localized chiral AFM-skyrmionic spin textures (Fig.~\ref{fig:1}c) and  metastable AFM domain walls (see Supplementary Figure 1) emerge in the Cr overlayer. Besides isolated topological AFM solitons, we identify strikingly unusual interpenetrating AFM skyrmions, which are reminiscent of crossing rings (see schematic Fig.~\ref{fig:1}b), the building blocks of knot theory where topological concepts such as Brunnian links are a major concept~\cite{Wu_knot1992}. The latter has far reaching consequences in various fields of research, not only in mathematics or physics but extends to chemistry and biology. For instance,  the exciting and intriguing interchain process, known also as catenation, is paramount in carbon-, molecular-, protein- or DNA-based assemblies~\cite{Amabilino1995interlocked,Dabrowski-Tumanski2017,bates2005dna}.  
We discuss the mechanisms enforcing the stability of the unveiled interchained topological objects, their response to magnetic fields and the subtle dependence on the underlying magnetic textures hosted in PdFe bilayer. Our findings are of prime importance in promoting AFM localized entities as information carriers in future AFM spintronic devices.

\section*{Results}

\subsection{AFM skyrmions in CrPdFe/Ir(111) surface.}

PdFe deposited on Ir(111) surface hosts a homo-chiral spin spiral as a ground state~\cite{romming2013writing,dupe2014tailoring} emerging from the interplay of the Heisenberg exchange interactions and DMI. The latter is induced by the heavy Ir substrate, which has  a strong spin-orbit coupling. Upon application of a magnetic field, sub 10-nm FM skyrmions are formed~\cite{romming2013writing,dupe2014tailoring,Simon2014,romming2015field,crum2015perpendicular,Bouhassoune2021,fernandes2022}. After deposition of the Cr overlayer, the magnetic interactions characterizing Fe are strongly modified (see comparison plotted in Supplementary Figure 2) due to changes induced in the electronic structure (Supplementary Figure 3). The Heisenberg exchange interaction among Fe nearest neighbors (n.n.) reduces by 5.5 meV (a decrease of 33\%). This enhances the non-collinear magnetic behavior of Fe, which leads to FM skyrmions even without the application of a magnetic field (see Supplementary Figure 4).
The n.n. Cr atoms couple strongly antiferromagnetically (-51.93 meV), which along with the antiferromagnetic coupling of the second n.n. (-6.69 meV) favors the N\'eel state. The subtle competition with the  ferromagnetic exchange interactions of the third n.n. (5.32 meV) stabilizes the RW-AFM state independently from the AFM interaction with the Fe substrate (the detailed magnetic interactions are shown in Supplementary Figure 2).
As illustrated in Fig.~\ref{fig:1}c, the RW-AFM configuration is characterized by parallel magnetic moments along a close-packed atomic row, with antiparallel alignment between adjacent rows. Due to the hexagonal symmetry of the atomic lattice, the AFM rows can be rotated in three symmetrically equivalent directions. We note that the moments point out-of-plane due to the magnetic anisotropy energy (0.5 meV per magnetic atom).

The DM interactions among Cr atoms  arise due to the  broken inversion symmetry and is mainly induced by the underlying Pd atoms hosting a large spin-orbit coupling. The n.n. Cr DMI (1.13 meV) is of the same chiral nature and order of magnitude as that of Fe atoms (1.56 meV), which gives rise to the chiral non-collinear behavior illustrated in Fig.~\ref{fig:1}c. We note that the solitons are only observed if Cr magnetic interactions beyond the n.n. are incorporated, which signals the significance of  the long-range coupling in stabilizing the observed textures. Since the Heisenberg exchange interaction among the Cr atoms is much larger than that of Fe, the AFM solitons are bigger, about a factor of three larger than the FM skyrmions found in Fe.

While the RW-AFM state is defined by two sublattices, the different AFM skyrmions, isolated or overlapped, can be decomposed into interpenetrating FM skyrmions living in 4 sublattices illustrated in Fig.~\ref{fig:1}d and denoted as L1, L2, L3 and L4. In the RW-AFM phase, L1 and L4 are equivalent and likewise for L2 and L3. It is evident that the moments in L1 and L4 are antiparallel to the ones in L2 and L3. Taking a closer look at the isolated AFM magnetic texture, one can dismantle it into two FM skyrmions with opposite topological charges anchored in the distinct antiparallel FM sublattices L1 and L2, while L3 and L4 carry rather collinear magnetization (Fig.~\ref{fig:1}e). In the case of the overlapped AFM skyrmions, however, no sublattice remains in the collinear state. As an example, the dimer  consists of two couples of antiferromagnetically aligned skyrmions, each being  embedded in one of the four sublattices (Fig.~\ref{fig:1}f).

Our study reveals that in contrast to interlinked magnetic textures, single AFM skyrmions are significantly sensitive to the magnetic environment hosted by the underlying PdFe bilayer. For brevity, we focus in the next sections on the single and two-overlapping AFM skyrmions and address  the mechanisms dictating their stability.

\subsection{Stabilization mechanism of the overlapping AFM skyrmions.}

The formation of overlapped solitons is an unusual phenomenon since FM skyrmions repel each other. It results from competing interactions among the skyrmions living in the different sublattices, which finds origin in the natural AFM coupling between the n.n. magnetic moments.
Depending on the hosting sublattice (L1 to L4), the four skyrmions shown in Fig.~\ref{fig:1}f experience attraction or repulsion.\ 
The sublattices are chosen such that nearest neighbours within a sublattice are third nearest neighbours in the overall system. This choice leads to the exchange coupling preferring the parallel alignment of spins within a given sublattice. When looking at any sublattice in isolation, this effective ferromagnetic-like exchange interaction enables the existence of skyrmions in a collinear background. In the overall system, however, pairs of sublattices interact via the first and second nearest-neighbour exchange interactions, which prefers anti-parallel spin alignments. Therefore, the exchange interaction between skyrmions formed at sublattices with a parallel background, such as (L1, L4) and (L2, L3), and denoted in the following as skyrmion-skyrmion homo-interactions, are repulsive as usually experienced by FM skyrmions.  In contrast, and for the same reasons, interaction between skyrmions in sublattices with oppositely oriented background spins, denoted as hetero-interactions, are attractive as it is for (L1, L2), (L2, L4), (L3, L4) and (L1, L3). Clearly, the set of possible hetero-interactions, enforced by the attractive nature induced by the DMI, outnumbers the homo ones. The interchained AFM skyrmion is simply the superposition of the sublattice skyrmions at the equilibrium distance, here 2.58 nm between the two AFM skyrmions, where both interactions (attraction and repulsion) are equal.

To substantiate the proposed mechanism, we quantify the skyrmion-skyrmion interaction.  We simplify the analysis by neglecting the Cr-Fe magnetic interactions, which puts aside the impact of the rich non-collinear magnetic behavior hosted by the PdFe bilayer. In this case, single AFM skyrmions disappear and only the overlapping ones are observed. We take the skyrmion dimer illustrated in Fig.~\ref{fig:2}a and proceed to a rigid shift of the lower AFM skyrmion while pinning the upper one at the equilibrium position. We extract the skyrmion-skyrmion interaction map as a function of distance, as shown in  Fig.~\ref{fig:2}b, which clearly demonstrates that as soon as the AFM skyrmions are pulled away from each other, the energy of the system increases. Note that within this procedure, the sublattice interactions (L1, L2) and (L3, L4) do not contribute to the plots since they are assigned to each of the AFM skyrmions moved apart from each other. Two minima are identified along a single direction as favored by the symmetry reduction due to the AFM arrangement of the magnetic moments in which the skyrmions are created. Indeed, one notices in Fig.~\ref{fig:1}d that due to the sublattice decomposition symmetry operations are reduced to C$_2$, i.e. rotation by $180^\circ$, while mirror symmetries, for example, originally present in the fcc(111) lattice are broken.
Fig.~\ref{fig:2}c and d depict the skyrmion-skyrmion interaction, which hosts either one or two minima, as a function of distance along two directions indicated by the dashed lines, blue and black, in Fig.~\ref{fig:2}a. 
The two minima found along the blue line should be degenerate and correspond to the swapping of the two AFM skyrmions. The breaking of degeneracy is an artifact of the rigid shift assumed in the simulations, which can be corrected by allowing the moments to relax (see red circle in Fig.~\ref{fig:2}c). The maximum of repulsion is realized  when the two AFM skyrmions perfectly overlap (see inset).

The interaction profile shown in Fig.~\ref{fig:2}d is decomposed into two contributions: the skyrmion-skyrmion homo- and hetero-interactions, which we plot in Fig.~\ref{fig:2}e and f respectively. The data clearly reveals the strong repulsive nature of the homo-interaction mediated by the Heisenberg exchange, which competes with the attractive hetero-interaction driven by both the Heisenberg exchange coupling and DMI. The latter skyrmion-skyrmion interaction is strong enough to impose the unusual compromise of having strongly overlapping solitons.

\subsection{Impact of magnetic field.}

Prior to discussing  stability aspects pertaining to the single AFM skyrmion in detail, we apply a magnetic field perpendicular to the substrate and disclose  pivotal ingredients for the formation of the isolated solitons. In general, the reaction of FM and AFM skyrmions to an external magnetic field is expected to be deeply different. When applied along the direction of the background magnetization, FM skyrmions reduce in size while recent predictions expect a size expansion of AFM skyrmions~\cite{bessarab2019stability,potkina2020skyrmions,rosales2015three}, thereby enhancing  their stability.  

To inspect the response of AFM skyrmions to a magnetic field perpendicular to the substrate, we first remove, as done in the previous section, the Cr-Fe interaction  since it gives rise to a non-homogeneous and strong effective exchange field. In this particular case, we can only explore the case of interchained AFM skyrmions. As illustrated in Fig.~\ref{fig:3}a, the size of each of the sublattice skyrmions, which together form the AFM skyrmion dimer, increases with an increasing magnetic field. The type of the hosting sublattice, with the magnetization being parallel or antiparallel to the applied field, seems important in shaping the skyrmions dimension.
Strikingly, and in strong contrast to what is known for FM skyrmions,  the AFM skyrmions, single and  multimers, were found to be stable up to extremely large magnetic fields. Although the assumed fields are unrealistic in the lab, they can be emulated by the exchange field induced by the underlying magnetic substrate. Indeed, the magnetic interaction between Cr and its nearest neighboring Fe atoms, carrying each a spin moment of 2.51 $\mu_B$, reaches -3.05 meV, which translates to an effective field of about 21 T. At this value, the average skyrmion radius is about 1.6 nm, which is 30\% smaller than the one found once the Cr-Fe magnetic coupling is enabled (see Fig.~\ref{fig:3}b). 

We note that since the skyrmions are  not circular in shape, their radius  is defined as the average distance between the skyrmion's center and the position where the spin moments lie in-plane. The significant size difference is induced by the strong inhomogenous exchange field emanating from the Fe sub-layer, which can host spirals, skyrmions and antiskyrmions. 

If the  Cr-Fe interaction is included, the size dependence changes completely. Instead of the rather monotonic increase with the field, the size of the skyrmion is barely affected until reaching about 50 T, which is accompanied by substantial miniaturization of the AFM skyrmions. Here, a phase transition occurs in Fe, which initially hosts spin spirals that turn into FM skyrmions (see Supplementary Figure 5). After being squeezed down to an average radius of 1.48 nm at  140 T, the size expansion observed without the Cr-Fe interaction is recovered because the substrate magnetization is fully homogeneous and parallel to the Zeeman field. Likewise, single AFM skyrmions, found only once the coupling to the substrate is enabled, react in a similar fashion to the field as depicted in Fig.~\ref{fig:3}c. The substantial difference, however, is that fields larger than 80 T destroy the AFM skyrmions due to the annihilation of the Fe FM skyrmions. This  highlights an enhanced sensitivity to the underlying magnetic environment and clearly demonstrates the robustness enabled by skyrmion interchaining.

\subsection{Stabilization mechanism for single AFM skyrmions.}

We learned that single AFM skyrmions can be deleted after application of an external magnetic field or by switching off the exchange coupling to the magnetic substrate. Both effects find their origin in the   magnetization behavior of the PdFe bilayer.   To explore the underlying correlation, we consider as an example the magnetic configuration obtained with a field of 70 T and delete one after the other the skyrmions and antiskyrmions found in Fe, then check whether the AFM skyrmion in Cr survives (see example in Fig.~\ref{fig:3}d). We notice that the AFM skyrmion  disappears by deleting the FM solitons located directly underneath or even a bit away.  Supplementary Figure 6 shows that when shifted across the lattice, the AFM skyrmion  disappears if fixed above a magnetically collinear Fe area.

We proceed in Fig.~\ref{fig:4} to an analysis of the Fe-Cr interaction pertaining to the lower-right snapshot presented in  Fig.~\ref{fig:3}d by separating the Heisenberg exchange contribution from that of DMI and plotting the corresponding site-dependent heat maps of these two contributions  for each sublattice. Here, we consider as reference energy that of the RW-AFM collinear state surrounding the non-collinear states in Fig.~\ref{fig:3}d. The building-blocks of the AFM skyrmion are shown in Fig.~\ref{fig:4}a and d, where one can recognize the underlying Fe FM skyrmions in the background. The latter are more distinguishable in the sublattices free from the AFM skyrmion as illustrated in Figs.~\ref{fig:4}g and j. The order of magnitude of the interactions clearly indicates that the DMI plays a minor role and that one can basically neglect the interactions arising in the skyrmion-free sublattices, namely L3 and L4. It is the Heisenberg exchange interaction emerging in the sublattices L1 and L2 that dictates the overall stability of the AFM skyrmion.

In L2, the core of the magnetization of the Cr FM skyrmion points along the same direction  as that of the underlying Fe atoms, which obviously is disfavored by the AFM coupling between Cr and Fe (-3.05 meV for nearest neighbors). This induces the red exchange area surrounding the core of the AFM skyrmion (black circle in Fig.~\ref{fig:4}e), which is nevertheless  sputtered with blue spots induced by the magnetization of the core of the Fe FM skyrmions pointing in the direction opposite to that of the Cr moments in L2. The latter is a mechanism reducing the instability of the Cr skyrmion. Overall, the total energy cost in having the Cr skyrmion in L2 reaches +693.7 meV and is compensated by the exchange energy of -712.4 meV generated by the one living in sublattice L1. Here, the scenario is completely reversed since the core of the Cr skyrmion has its magnetization pointing in the opposite direction than that of the neighboring Fe atoms and therefore the large negative blue area with the surrounding area being  sputtered by the Fe skyrmions, similar to the observation made in L2 (see Fig.~\ref{fig:4}d). Overall, the Cr AFM  skyrmion arranges its building blocks such that the energy is lowered by the skyrmion anchored in sublattice L1. Here, the details of the non-collinear magnetic textures hosted by Fe play a primary role in offering the right balance to enable stabilization. This explains the sensitivity of the single AFM skyrmion to the number and location of the underlying FM Fe skyrmions. Removing non-collinearity in Fe makes both building blocks of the AFM skyrmion equivalent without any gain in energy from the Cr-Fe interaction, which facilitates the annihilation of the Cr skyrmion.

\subsection{Phase diagrams.}

It is enlightening to explore the phase diagrams of the AFM skyrmions as function of the underlying magnetic interactions. The latter are multiplied by a factor renormalizing the initial parameters. In Fig.~\ref{fig:5} we illustrate the impact of DMI vector's  (\textbf{D}) magnitude, Heisenberg exchange $J$ and anisotropy $K$ on the formation of various phases including the one hosting double overlapped AFM skyrmions. For simplicity, we consider the case where the interaction between Cr and the underlying Fe layer is switched off. A color code is amended to follow the changes induced on the distance between the AFM skyrmions. From this study, we learn that in contrast to the DMI, which tends to increase the size of the structures, $J$ and $K$ tend to miniaturize the skyrmions, ultimately favoring  their annihilation. The phase hosting AFM skyrmions is sandwiched between the RW-AFM state and a phase hosting stripe domains. It is convenient to analyse the unveiled overall behavior in terms of the impact of DMI. The latter protects the AFM skyrmions structure from shrinking, similarly to FM skyrmions~\cite{sampaio2013nucleation,rohart2013skyrmion}. So for small values of DMI compared to $J$ in Fig. ~\ref{fig:5}a, or compared to $K$ in Fig. ~\ref{fig:5}b, the AFM skyrmions shrink and disappear. In contrast,  large values of the DMI increase the size of the skyrmions till reaching a regime where stripe domains are formed. Within the phase hosting AFM skyrmions, increasing $J$ or $K$ results in smaller skyrmions.

\subsection{Thermal stability with Geodesic nudged elastic band (GNEB) method. }

So far we have demonstrated that the interlinked AFM skyrmion multimers can indeed exist as local minima of the energy expression given by the Heisenberg Hamiltonian (Eq.~\ref{eq.Heisenberg}). Another important question, however, is the stability of these structures against thermal excitations.  
Answering this question requires knowledge about how deep or shallow these energy minima are, which can be quantified as a minimal energy barrier that the system has to overcome in order to escape a minimum, keeping in mind that the N\'eel temperature of the RW-AFM ground state is $\approx$ 310 K as obtained from our Monte Carlo simulations ~\cite{hinzke1999monte,nowak2001thermally,muller2019spirit}. To investigate this issue, we systematically carried out a series of Geodesic nudged elastic band (GNEB) simulations~\cite{bessarab2015method,muller2018duplication,muller2019spirit}  for AFM multimers, containing initially 10 interchained skyrmions, then calculating the energy barrier needed to annihilate one AFM skyrmion at a time as depicted in Fig. ~\ref{fig:55}a, showing the successive magnetic states between which, the energy barrier has been calculated. Note that deleting one of the AFM skyrmions forming the dimer leads to the RW-AFM state.  The energy barrier is given by the energy difference between the $n^{th}$ AFM skyrmions state local minimum (hosting $n$ AFM interchained skyrmions) and the relevant saddle point located on the minimum energy path connecting the initial state with the $(n-1)^{th}$ AFM skyrmions state. The  energy barrier increases from about 8 meV ($\approx$ 90 K)  for the double interchained AFM skyrmions to 13 meV  ($\approx$ 150 K) for three interchained ones, reaching a saturation value of $\approx$ 18.5 meV ($\approx$ 214 K)  for chains containing more than five AFM skyrmions, see Fig.~\ref{fig:55}b. Hence, increasing the number of interchained skyrmions enhances their stability, which is further amplified when enabling the interaction with the PdFe substrate.
Instead of 8 meV pertaining to the free skyrmion dimer, the barrier reaches 45.7 meV ($\approx$ 530 K) owing to the interaction with the underlying substrate while the single AFM skyrmion experiences a barrier of 10 meV ($\approx$ 113 K). Thus, the exchange field emanating from the PdFe substrate promotes the use of interchained AFM skyrmions in room temperature applications. By analysing how the different interactions contribute to the barrier, we identified the DMI as a key parameter for  the thermal stability of  the interchained AFM skyrmions. For example, in the case of free double interchained AFM skyrmions, the Heisenberg exchange interactions contribution is -87 meV, the magnetic anisotropy contribution is -150 meV while the DMI provides a barrier of 245 meV. Interestingly and as expected, it is the magnetic exchange interaction between Cr and Fe that is mainly responsible for the thermal stability of the single AFM skyrmion.
\\

\section*{Discussion}

Following a two-pronged approach based on first-principles simulations combined with atomistic spin dynamics, we identify a thin film that can host intrinsic, i.e. non-synthetic, AFM skyrmions at zero magnetic field. 
A Cr monolayer deposited on a substrate known to host FM skyrmions, PdFe/Ir(111), offers the right AFM interface combination enabling the emergence of a rich set of AFM topological solitons. Owing to the AFM nature of Cr, its  ground state is RW-AFM as induced by magnetic interactions beyond nearest neighbors. We strikingly discovered interchained AFM skyrmions, which can be cut and decoupled into isolated solitons via  the inhomogenous exchange field emanating from PdFe bilayer. Interestingly, interchaining enhances their stability, which is largely amplified by the exchange field emanating from the substrate.
The intra-overlayer skyrmion-skyrmion interaction favors the important overlap of the AFM skyrmions, and makes them robust against the rich magnetic nature of the PdFe substrate. In contrast, the single AFM skyrmion annihilates if positioned on a homogeneously magnetized substrate and it is only through the presence of various spin-textures such as spin spirals and multiples of skyrmions or antiskyrmions that one of the building-blocks of the AFM skyrmion can lower the energy enough to enable stability.

Since the experimental observation of intrinsic AFM skyrmions has so far been elusive at interfaces, our predictions open the door for their realization in well defined materials and offer the opportunity to explore them in thin film geometries. The robustness of the interchained skyrmions qualify them as ideal particles for room temperature racetrack memory devices to be driven with currents while avoiding the skyrmion Hall effect. Preliminary work indicates that AFM skyrmions
move faster than their FM partners when reacting to an applied current, with an intriguing behavior induced by the  non-collinear exchange field emanating from the substrate. If the latter is collinear or non-magnetic, the AFM textures are predicted to be quasi-free from the skyrmion Hall effect. The ability to control the substrate magnetization makes the system, we studied,  a rich playground to design and tune the highly sensitive single AFM skyrmion living in the overlayer. We envision patterning the ferromagnetic surface with regions hosting different  magnetic textures, being trivial, such as the ferromagnetic regions, or topological, to define areas where the AFM skyrmion can be confined or driven within specific paths. We envisage a rich and complex response to in-plane currents, which could move the underlying FM solitons along specific directions subjected to the Magnus effect. The latter would affect  the response of the overlaying isolated AFM skyrmions in a non-trivial fashion. 

We anticipate that the proposed material and the FM-substrate on which AFM films can be deposited offer an ideal platform to explore the physics of AFM skyrmions. Noticing already that different overlapped AFM skyrmions can co-exist establishes a novel set of multi-soliton objects worth exploring in future studies. Besides their fundamental importance enhanced by the  potential parallel with topological concepts known in knot theory, such unusual AFM localized entities might become  exciting and useful constituents of future nanotechnology devices resting on non-collinear spin-textures and the emerging field of antiferromagnetic spintronics.

\begin{methods}

\subsection{First-principles calculations.}

The relaxation parameters were obtained using the Quantum-Espresso computational package\cite{giannozzi2009quantum}. The projector augmented wave pseudo potentials from the PS Library \cite{dal2014pseudopotentials} and a $28\times28\times1$  k-point grid were used for the calculations. The Cr, Pd, Fe and Ir interface layer were fcc-stacked along the [111] direction and relaxed by 4\%, 5.8\%, 8.1\% and -1\% with respect to the Ir ideal interlayer distance, respectively. Positive (negative) numbers refer to atomic relaxations towards (outward of) the Ir surface.

The electronic structure and magnetic properties were simulated using all electron full potential scalar relativistic Koringa -Kohn-Rostoker (KKR) Green function method \cite{Papanikolaou2002,Bauer2014} in the local spin density approximation. The slab contains 30 layers  (3 vacuum + 1 Cr + 1 Pd + 1 Fe + 20 Ir + 4 vacuum). The momentum expansion of the Green function was truncated at $\ell_{\text{max}} = 3$. The self-consistent calculations were performed with a k-mesh of $30\times30$ points and the energy contour contained 23 
complex energy points in the upper complex plane with 9 Matsubara poles. The Heisenberg exchange interactions and DM vectors were extracted using the infinitesimal rotation method~\cite{Liechtenstein1987,Ebert2009} with a k-mesh of a $200\times200$.

 \subsection{Hamiltonian Model and atomistic spin dynamics.}
 In our study, we consider a two dimensional  Heisenberg model on a triangular lattice, equipped with Heisenberg exchange coupling, DMI, the magnetic anisotropy energy, and Zeeman term. All  parameters were obtained from ab-initio. The energy functional reads as follows:

 \begin{equation}
   H = H_\text{Exc} + H_\text{ DMI} + H_\text{Ani} + H_\text{Zeem},
   \label{eq.Heisenberg}
\end{equation}
with:
\begin{equation*}
 H_\text{Exc}= -\sum\limits_{<i,j>} J^\text{Cr-Cr}_{ij}\; \textbf{S}_{i}\cdot \textbf{S}_{j}       -\sum\limits_{<i,j>} J^\text{Fe-Cr}_{ij}\; \textbf{S}_{i}\cdot \textbf{S}_{j}   -\sum\limits_{<i,j>} J^\text{Fe-Fe}_{ij}\;\textbf{S}_{i}\cdot \textbf{S}_{j},
\end{equation*}
\begin{equation*}
  H_\text{DMI}= \sum\limits_{<i,j>}\textbf{D}^\text{Cr-Cr}_{ij}\cdot [\textbf{S}_{i}\times \textbf{S}_{j}]+\sum\limits_{<i,j>}\textbf{D}^\text{Fe-Cr}_{ij}\cdot [\textbf{S}_{i}\times \textbf{S}_{j}]+\sum\limits_{<i,j>}\textbf{D}^\text{Fe-Fe}_{ij}\cdot [\textbf{S}_{i}\times \textbf{S}_{j}],  
 \end{equation*}
 \begin{equation*}
 H_\text{Ani}=- K^\text{Cr}\sum\limits_{i}  (S_i ^z)^2 - K^\text{Fe}\sum\limits_{i}  (S_i ^z)^2,
 \end{equation*}
 \begin{equation*}
H_\text{Zeem} =- \sum\limits_{i} h_i S_i^z,  
\end{equation*}
  where $i$ and $j$ are site indices carrying each magnetic moments.  $\textbf{S}$ is a unit vector of the magnetic moment. $J^\text{X-Y}_{ij}$ is the Heisenberge exchange coupling strength, being $<$ 0 for AFM interaction, between an X atom on site $i$ and a Y atom on site $j$. A similar notation is adopted for the DMI vector $\textbf{D}$ and the magnetic anisotropy energy $K$ (0.5 meV per magnetic atom). The latter favors the out-of-plane orientation of the magnetization, and $h_i= \mu_i B$  describes the Zeeman coupling to the atomic spin moment $\mu$ at site $i$ assuming an out-of-plane field.

To explore the magnetic properties and emerging complex states we utilize the Landau-Lifshitz-equation (LLG) as implemented in the Spirit code~\cite{muller2019spirit}. We assumed periodic boundary conditions to model the extended two-dimensional system with cells containing $100^2$, $200^2$, $300^2$, $400^2$ sites. 

\subsection{Data availability}
The data needed to evaluate the conclusions in the paper are present in the paper and the Supplementary Information.

\subsection{Code availability} We used the following codes:
Quantum ESPRESSO,  SPIRIT can be found at \url{https://github.com/spirit-code/spirit}, and 
the KKR code is a rather complex ab-initio DFT-based code, which is in general impossible to use without proper training on the theory behind it and on the practical utilization of the code. We are happy to provide the latter code upon request.

\end{methods}

\begin{addendum}

\item We thank Markus Hoffmann for fruitful discussions. This work was supported by the Federal Ministry of Education and Research of Germany
in the framework of the Palestinian-German Science Bridge (BMBF grant number
01DH16027) and the Deutsche For\-schungs\-gemeinschaft (DFG) through SPP 2137 ``Skyrmionics'' (Projects LO 1659/8-1, BL 444/16-2). 
The authors gratefully acknowledge
the computing time granted through JARA on the supercomputer JURECA 
at Forschungszentrum Jülich.

\item[Author contributions] S.L. initiated, designed and supervised the project. A.A. performed the simulations with support and supervision from I.L.F., S.Br. and M.S.  A.A., I.L.F., S.Br., M.S., M.A., S.Bl. and S.L. discussed the results. A.A. and S.L. wrote the manuscript to which all co-authors contributed.

\item[Competing Interests] The authors declare no competing interests.

\item[Correspondence] Correspondence and requests for materials should be addressed to A.A. (email: a.aldarawsheh@fz-juelich.de) or to S.L. (email: s.lounis@fz-juelich.de).

\end{addendum}
\begin{figure*}
\centering
\includegraphics[width=1.0\linewidth]{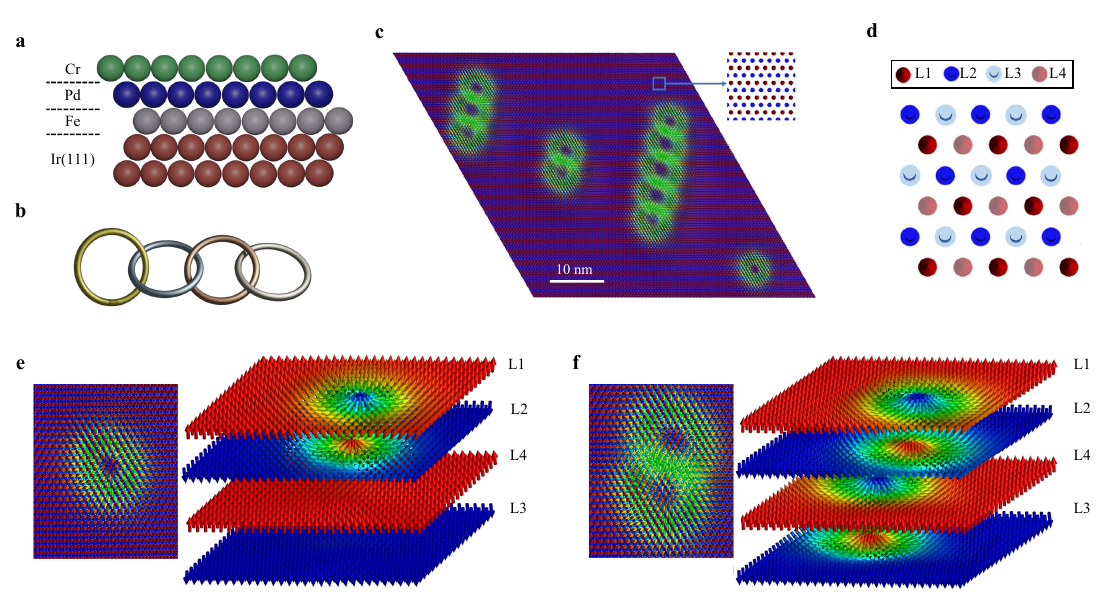}
\caption{\textbf{Interchained AFM skyrmions in CrPdFe trilayer on Ir(111).}  \textbf{a} Schematic representation of the investigated trilayer deposited on Ir(111) following fcc stacking.  \textbf{b} The interchaining of skyrmions is reminiscent of interpenetrating rings, which realize topologically protected phases. \textbf{c} The ground state being RW-AFM  (see inset) can host AFM skyrmions that can be isolated or interlinked to form multimers of skyrmions, here we show examples ranging from dimers to pentamers. The AFM skyrmions can be decomposed into FM skyrmions living in sublattices illustrated in \textbf{d}. In case of the single AFM skyrmion, two of the sublattices, L1 and L2, are occupied by the FM skyrmions shown in \textbf{e}.  L3 and L4 host quasi-collinear AFM spins in contrast to the FM skyrmions emerging in the case of the double AFM skyrmion presented in \textbf{f}. Note that the separation of sublattices L1, L2, L3 and L4 shown in \textbf{e} and \textbf{f} is only done for illustration.}
\label{fig:1}

\end{figure*}

\begin{figure*}
\centering
\includegraphics[width=1.0\linewidth]{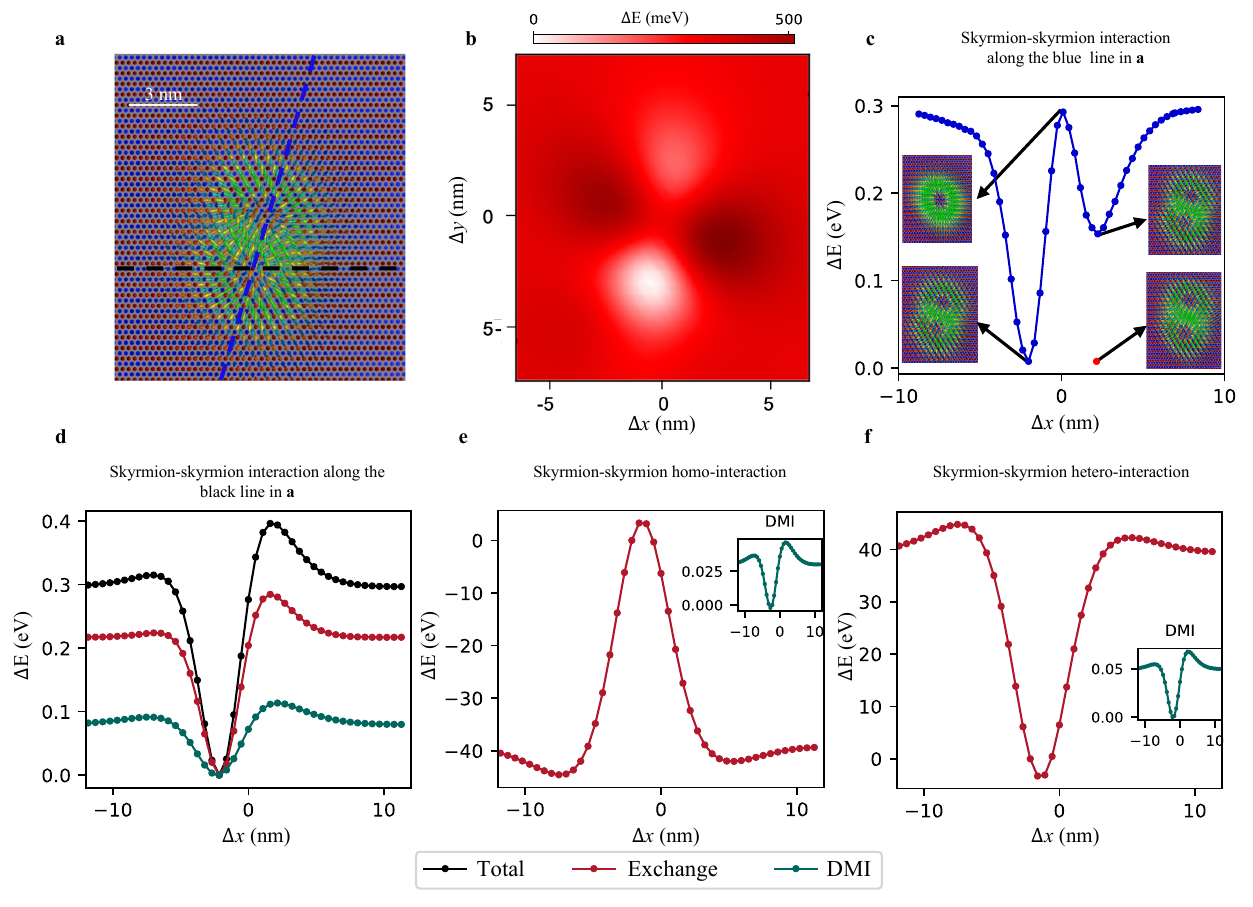}
\caption{\textbf{Energetics of two interchained AFM skyrmions.}
\textbf{a} Two overlapping AFM skyrmions decoupled from the PdFe bilayer with black and blue line representing two examples of paths along which the lower skyrmion is rigid-shifted with respect to the upper one, which is pinned. \textbf{b} Two-dimensional map of the total energy difference with respect to the  magnetic state shown in \textbf{a} as a function of the  distance between the skyrmion centers. \textbf{c} Energy profile along the blue line shown in \textbf{a}. A double minimum is found once the skyrmions swap their positions and become truly degenerate once the rigidity of the spin state is removed (see the red circle). \textbf{d} The Heisenberg exchange is the most prominent contribution to the skyrmion stabilization, as shown along the path hosting a single minimum.  The total skyrmion-skyrmion repulsive homo-interaction is dominated by the attractive hetero-interaction,  red curves in \textbf{e} and \textbf{f} respectively. The DMI contribution, shown in insets, is smaller and  sublattice independent. It favors the overlap of AFM skyrmions.}
\label{fig:2}
\end{figure*}

\begin{figure*}
\centering
\includegraphics[width=1.0\linewidth]{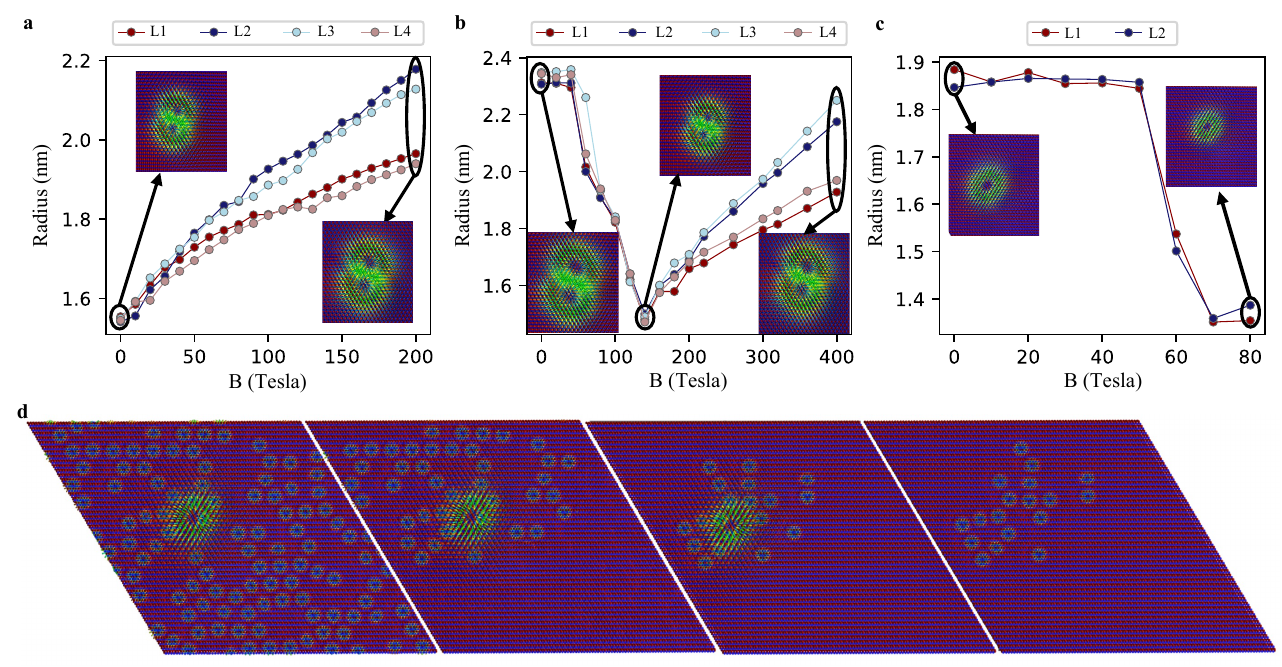}
\caption{\textbf{Impact of magnetic field of AFM skyrmion radius.} Radius of the sublattice FM skyrmions for two-interchained AFM skyrmions \textbf{a} decoupled from and \textbf{b} coupled to the Fe magnetization. \textbf{c} A single case is shown for the isolated AFM skyrmion since it disappears without the inhomogeneous magnetic field emerging from the substrate. Examples of snapshots of the AFM skyrmions  are illustrated as insets of the different figures. In Fe, the amount of FM skyrmions and antiskyrmions increases once applying a magnetic field, which erases the ground state spin-spiral. The coupling to the Fe magnetization affects  the evolution of the AFM skyrmions as function of the magnetic field dramatically. \textbf{d} depicts the dependence of the AFM skyrmion under a magnetic field of 70 Tesla on the surrounding magnetic environment, by sequentially deleting one FM skyrmion or antiskyrmion  in the Fe layer and relaxing the spin structure. At some point, removing any of the single FM skyrmions in the lower left of \textbf{d} annihilates the AFM skyrmion.}
\label{fig:3}
\end{figure*}
\begin{figure*}
\centering
\includegraphics[width=1.0\linewidth]{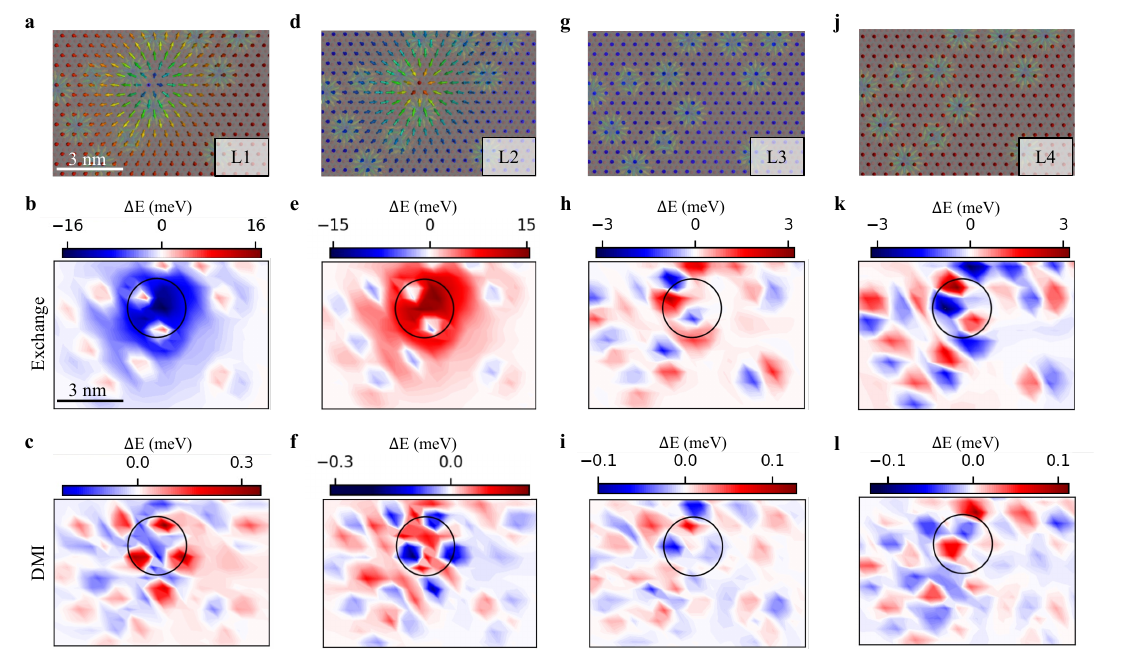}
\caption{\textbf{Interaction map of the single AFM skyrmion with the magnetic substrate.} In the first row of figures, sublattice decomposition of Cr skyrmion including the underlying Fe skyrmions shown in four columns~\textbf{a}, \textbf{d}, \textbf{g} and \textbf{j}, corresponding respectively to L1, L2, L3 and L4. The AFM skyrmion is made of two FM skyrmions hosted by sublattices L1 and L2. In Fe, FM skyrmions and antiskyrmions can be found in all four lattices. 
The second row (\textbf{b}, \textbf{e}, \textbf{h} and \textbf{k})  illustrates the sublattice dependent two dimensional Heisenberg exchange energy map corresponding to the areas plotted in the first row, followed by the third row (\textbf{c}, \textbf{f}, \textbf{i} and \textbf{l}) corresponding to DMI. Note that the energy difference $\Delta E$ is defined with respect to the RW-AFM background.}
\label{fig:4}
\end{figure*}
\begin{figure*}
\centering
\includegraphics[width=1.0\linewidth]{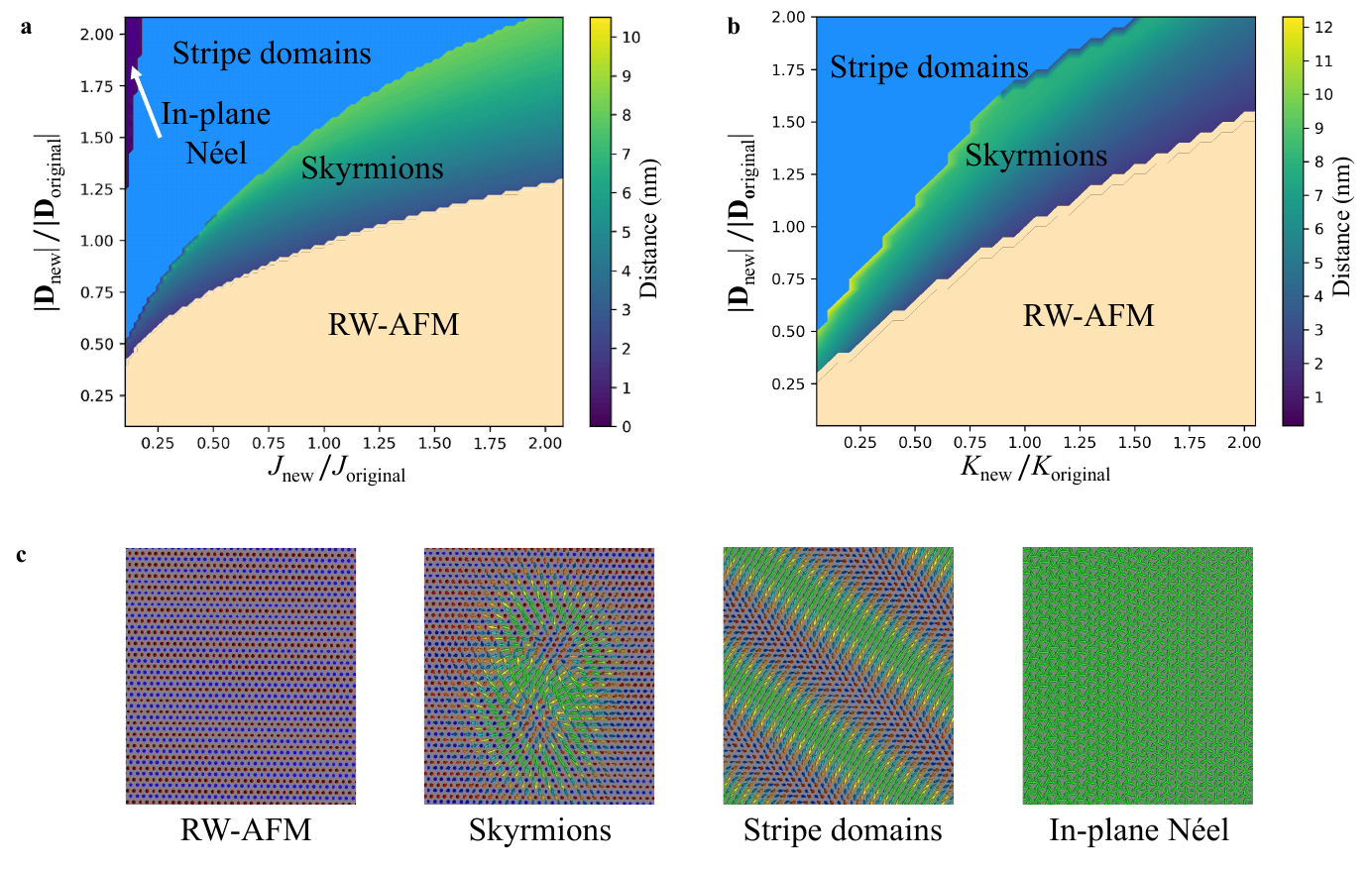}
	\caption{\textbf{ 
	 Phase diagrams of the free double interchained AFM skyrmions.} \textbf{a} Phase diagram obtained by fixing the magnetic anisotropy energy $K$ while changing the set of DMI and   Heisenberg exchange interaction $J$, or \textbf{b} by fixing $J$ while modifying $K$ and DMI. The color gradient pertaining to the skyrmion phase indicates the distance between two AFM skyrmions. \textbf{c} Illustration of the states
       shown in the phase diagrams. Note that an in-plane N\'eel state is predicted for large DMI and small $J$.}
\label{fig:5}
\end{figure*}
\begin{figure*}
\centering
\includegraphics[width=1.0\linewidth]{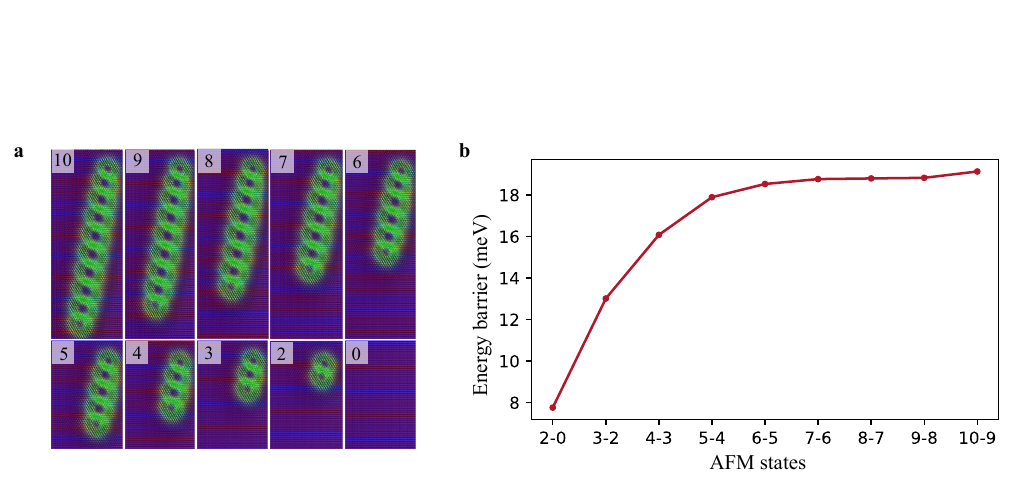}
\caption{\textbf{Energy barriers for chains of free interchained AFM skyrmions.}
 \textbf{a} Snapshots of the explored skyrmion chains. \textbf{b} The energy barrier obtained with GNEB simulations for deleting a single AFM skyrmion from the lower edge of the free (not interacting with PdFe) chains. }
\label{fig:55}
\end{figure*}
\end{document}